\documentclass[reprint,superscriptaddress,amsmath,amssymb,aps,prl]{revtex4-2}
\usepackage[utf8]{inputenc} %
\usepackage{graphicx}
\usepackage{dcolumn}
\usepackage{bm}
\usepackage{amsmath}
\usepackage{rotating}
\usepackage{enumerate}
\usepackage{xcolor}

\usepackage{array,multirow,graphicx}
\usepackage{float}
\usepackage{hyperref}
\hypersetup{colorlinks,allcolors=black}
\makeatletter
\renewcommand*{\doi}[1]{DOI: \href{http://dx.doi.org/#1}{#1}}
\begin{document}

\title{Transferable classical force field for pure and mixed metal halide perovskites parameterized from first principles}

\author{Juan Antonio Seijas-Bellido}
\affiliation{\'Area de Qu\'imica F\'isica, Universidad Pablo de Olavide, 41013 Seville, Spain}

\author{Bipasa Samanta}
\affiliation{Department  of  Materials  Science  and  Engineering,  Technion--Israel  Institute  of Technology, Haifa 3200003, Israel}

\author{Karen Valadez-Villalobos}
\affiliation{\'Area de Qu\'imica F\'isica, Universidad Pablo de Olavide, 41013 Seville, Spain}

\author{Juan Jes\'us Gallardo}
\affiliation{Departamento de Qu\'imica F\'isica, Facultad de Ciencias, Universidad de C\'adiz, E-11510 Puerto Real, C\'adiz, Spain}

\author{Javier Navas}
\affiliation{Departamento de Qu\'imica F\'isica, Facultad de Ciencias, Universidad de C\'adiz, E-11510 Puerto Real, C\'adiz, Spain}

\author{Salvador R. G. Balestra}
\affiliation{\'Area de Qu\'imica F\'isica, Universidad Pablo de Olavide, 41013 Seville, Spain}
\affiliation{Instituto de Ciencia de Materiales de Madrid, Consejo Superior de Investigaciones Científicas (ICMM-CSIC) c/Sor Juana Inés de la Cruz 3, Madrid 28049, Spain}

\author{Rafael Mar\'ia Madero-Castro}
\affiliation{\'Area de Qu\'imica F\'isica, Universidad Pablo de Olavide, 41013 Seville, Spain}

\author{Jose Manuel Vicent-Luna}
\affiliation{Materials Simulation and Modelling, Department of Applied Physics, Eindhoven University of Technology, P.O. Box 513, 5600MB, Eindhoven, The Netherlands}

\author{Shuxia Tao}
\affiliation{Materials Simulation and Modelling, Department of Applied Physics, Eindhoven University of Technology, P.O. Box 513, 5600MB, Eindhoven, The Netherlands}

\author{Maytal Caspary Toroker}
\affiliation{Department  of  Materials  Science  and  Engineering,  Technion--Israel  Institute  of Technology, Haifa 3200003, Israel}

\author{Juan Antonio Anta}
\affiliation{\'Area de Qu\'imica F\'isica, Universidad Pablo de Olavide, 41013 Seville, Spain}
\email{jaantmon@upo.es}

\date{\today}

\begin{abstract}
Many key features in photovoltaic perovskites occur in relatively long time scales and involve mixed compositions. This requires realistic but also numerically simple models. In this work we present a transferable classical force field to describe the mixed hybrid perovskite MA$_x$FA$_{1-x}$Pb(Br$_y$I$_{1-y}$)$_3$ for variable composition ($\forall x,y \in [0,1]$). The model includes Lennard-Jones and Buckingham potentials to describe the interactions between the atoms of the inorganic lattice and the organic molecule, and the AMBER model to describe intramolecular atomic interactions. Most of the parameters of the force field have been obtained by means of a genetic algorithm previously developed to parameterize the CsPb(Br$_x$I$_{1-x}$)$_3$ perovskite~\cite{D0TA03200J}. The algorithm finds the best parameter set that simultaneously fits the DFT energies obtained for several crystalline structures with moderate degrees of distortion with respect to the equilibrium configuration. The resulting model reproduces correctly the XRD patterns, the expansion of the lattice upon I/Br substitution and the thermal expansion coefficients. We use the model to run classical molecular dynamics simulations with up to 8600 atoms and simulation times of up to 40~ns. From the simulations we have extracted the ion diffusion coefficient of the pure and mixed perovskites, presenting for the first time these values obtained by a fully dynamical method using a transferable model fitted to first principles calculations. The values here reported can be considered as the theoretical upper limit for ion migration dynamics induced by halide vacancies in photovoltaic perovskite devices under operational conditions.

\end{abstract}

\maketitle

%
%

\section{Introduction}
Metal halide perovskites (MHP) have emerged as one of the most studied semiconductors due to their excellent optoelectronic properties~\cite{doi:10.1021/jz4020162,doi:10.1021/acs.accounts.5b00455}. This is evidenced by the rapid development of perovskite solar cells (PSCs) with a record certified photoconversion efficiency of 23.7\%, similar to those of silicon cells~\cite{OxfordPV}. Nonetheless, industrial application of PSCs is critically hampered by instability issues, including intrinsic, environmental, and operational factors. Instability is attributed to several chemical and dynamical processes that occur at very distinct time scales, like ionic rearrangements and physical and chemical interactions in the bulk and at interfaces with contact layers. These phenomena cause current-voltage hysteresis, and ultimately, device degradation.

\begin{figure}
\includegraphics[width=0.48\textwidth]{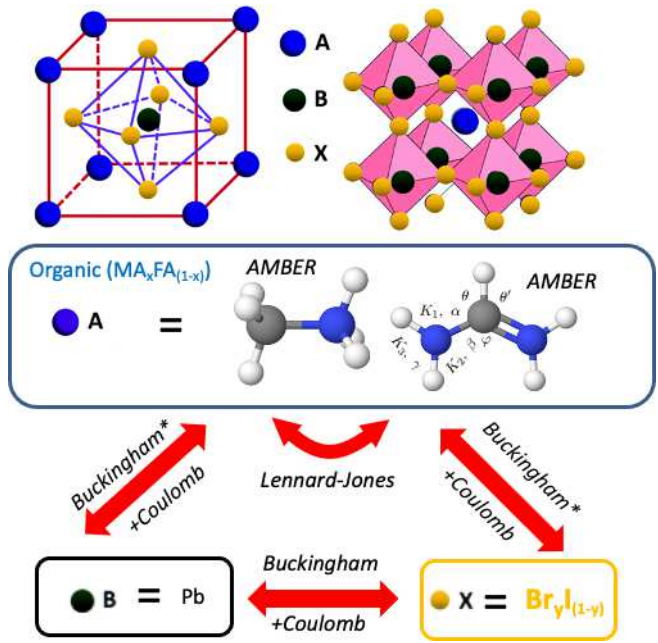}
\caption{Top: crystalline structure of metal halide perovskites. Middle: Methylammonium (MA, left) and formamidinium (FA, right) ions. The atoms included in the molecules are C (grey), N (blue) and H (white). Useful notation has been included in the FA molecule. (Figure created using https://molview.org.) Bottom: interactions considered between the different components of the perovskite: Buckingham, Lennard-Jones, Coulomb, AMBER (see text for more details). Note that a modified Buckingham potential is being used for the inorganic-organic interaction: ``Buckingham*" means that Lennard-Jones interactions were considered for the hydrogen atoms only.}
\label{fig:MA_FA_molecules}
\end{figure}

It has been proved that the inclusion of organic molecules as the A ion in the ABX$_3$ formula of the perovskites increases the efficiency and the stability of the cells. Promising results have been recently reported for FAPbI$_3$ solar cells, in which the efficiencies obtained reach values larger than 25\%~\cite{Min749,doi:10.1021/acsenergylett.9b02348,min2021}. However, the structural phase that allows for the absorption of photons and the creation of electron-hole pairs, the $\alpha$-black phase, is unstable, and rapidly decays to the $\delta$-yellow phase, useless for solar cell applications~\cite{doi:10.1021/acsenergylett.9b02348,doi:10.1021/acsenergylett.9b02450}. A possible solution for that is to mix different ions and molecules in the same active layer, like the caesium ion and the methylammonium (MA) and formamidinium (FA) molecules (Fig.~\ref{fig:MA_FA_molecules}), which significantly increases its stability without paying a high cost in efficiency.

The combination of A = Cs, MA, FA; B = Pb; X = Br, I is well known for researchers working in perovskite solar cells (PSCs). Many theoretical and experimental papers reported the structural parameters and the different properties of this kind of systems. Nevertheless, most of the theoretical studies have been performed using Density Functional Theory (DFT) calculations, which imply to work with a small number of atoms and allow for very short time dynamics simulations. Some phenomena that have been experimentally observed and described in these mixed supercells, such as ionic diffusion and segregation, cannot be modelled using first principles techniques. This is one of the reasons that leads us to think that a classical model for hybrid perovskites would be very useful, since a parameter such as the diffusion coefficient would be easily accessible through a classical molecular dynamics (CMD) simulation.

Two classical force fields (FF) developed in the University of Cagliari by Mattoni and coworkers for the MAPbBr$_3$~\cite{doi:10.1021/acs.jpcc.6b11298} and the MAPbI$_3$~\cite{Mattoni_2016} reproduce accurately the behavior of these two compounds. However, these models are not transferable, which means that they cannot describe mixed hybrid perovskites. With this motivation, some of the authors of this paper developed a FF that properly describes the properties and the behavior of the compositional set CsPb(Br$_x$I$_{1-x}$)$_3$ ($\forall x \in [0,1]$)~\cite{D0TA03200J}. To parameterize the mentioned FF, they developed a homemade genetic algorithm (GA). It uses the energies computed by an atomistic classical code to calculate the energies of different structures that can be compared with reference DFT results to achieve an optimal fitting of the parameters.

Following the same line of research, here we present a classical model potential, that is, a classical FF, capable of adequately describe the mixed hybrid perovskite MA$_x$FA$_{1-x}$Pb(Br$_y$I$_{1-y}$)$_3$, with $x$ and $y$ ranging between 0 and 1. We have checked that the model reproduces the lattice parameter of each pure compound as reported in the literature. We have also produced experimental and theoretical XRD patterns to prove that the structure is also well described. After that, by running molecular dynamics simulations with the fitted FF, we have predicted some results for different combinations of ions. For example, we have obtained the diffusion coefficients of the I and Br ions at realistic conditions of pressure, temperature and number of crystalline defects characteristic of these materials in a solar cell in operation.

The results obtained in this work compare very well with other theoretical and experimental data reported in the literature for crystalline perovskites. However, the results produced for the ionic diffusion coefficients cannot be directly extrapolated with the values that justify the low frequency behaviour of perovskite solar cell in operating conditions (hysteresis, impedance spectra, etc.). Still, they represent the expected value for a perfectly crystalline solid with ionic halide vacancies but with no other type of defects such as grain boundaries. Considering that they were actually derived from a classical FF fitted to DFT energies, the ion diffusion coefficients here reported can be considered as the upper theoretical limit for ion dynamics in perovskite solar cells. 

\section{The classical force field}
To describe the interaction of the C and N ions included in the MA and the FA and the inorganic lattice, i.e., Pb, I, Br ions, and between the ions of the inorganic lattice themselves, we have used the Buckingham potential, plus the Coulomb interaction 
\begin{align} \label{eq:Buck}
U(r_{ij})=A_{ij} \exp{\left(-\dfrac{r_{ij}}{\rho_{ij}}\right)} -\dfrac{c_{ij}}{r_{ij}^6}+\dfrac{q_iq_j}{4\pi \epsilon_0r_{ij}}.
\end{align}
It includes a repulsive term and an attractive term, and is usually chosen because the Pauli repulsion is well described by its exponential form. This potential model have also proved to work accurately for many materials, including other perovskites as mentioned before~\cite{D0TA03200J,doi:10.1021/acs.jpcc.6b11298,Mattoni_2016}. In this work, based on the relative weight of the attractive term with respect to the Coulomb part, we have set the "c" parameter in all Buckingham interactions to zero. See the Supporting Information for a more detailed justification of this choice (Figure S1).

To describe the non-bonded terms, which includes the interaction between atoms of different organic molecules, and all the interactions including hydrogen atoms, we have employed a Lennard-Jones potential plus the Coulomb interaction
\begin{align} \label{eq:LJ}
U(r_{ij})=4\epsilon_{ij}\left[ \left(\dfrac{\sigma_{ij}}{r_{ij}} \right)^{12}-\left(\dfrac{\sigma_{ij}}{r_{ij}} \right)^6 \right]+\dfrac{q_iq_j}{4\pi \epsilon_0r_{ij}}.
\end{align}
This potential also includes a repulsive and an attractive term, in a simpler way than the Buckingham potential. It have been used in many previous works to describe this type of interactions.~\cite{doi:10.1021/ja00124a002,jorgensen_development_1996,martin_transferable_1998,brooks_charmm_1983} 

Organic molecules (Fig.~\ref{fig:MA_FA_molecules}) are modeled by means of the AMBER model.~\cite{doi:10.1021/ja00124a002} This model maintains the shape of the molecule by including permanent harmonic bonds and angles, and also permanent dihedrals. The energy of the molecule is calculated via: 
\begin{align} \label{eq:AMBER}
U_{\text{bonded}}&=U_{\text{bonds}}+U_{\text{angles}}+U_{\text{dihedrals}}
\nonumber \\ &=\sum_{ij}K^b_{ij} \left(r_{ij}-r^0_{ij}\right)^2
\nonumber \\ &+\sum_{ijk}K^a_{ijk} \left(\theta_{ij}-\theta^0_{ij}\right)^2
\nonumber \\ &+\sum_{ijkl}K^d_{ijkl} \left(1+\cos(n_{ijkl}\phi_{ijkl}-\phi^0_{ijkl})\right).
\end{align}
This model have been used to describe many different organic molecules~\cite{https://doi.org/10.1002/jcc.20035}. It also allows for the possibility of distinguishing between different realizations of an atom of the same element. For instance, the parameters of the $sp^3$ C of the MA can be different than the parameters of the $sp^2$ C of the FA. Moreover, the parameters of the Ns depend of the number of ions they are bonded with. And the Hs bonded to a C or a N are also different. All these details will be further discussed in the parameterization section.

The ions in the organic molecules placed at a distance of 1 or 2 bonds (for example, the C-N in the FA and the N-N in the FA respectively) only interact by the bonded terms. That means that a pair of bonded atoms is only determined by the bond, angle and dihedral potentials. Ions placed at a distance of 3 bonds (for example a C-bonded hydrogen or a N-bonded hydrogen) feel the Lennard-Jones interaction and the Coulomb interaction, but not at its full strength. We set weighting coefficients of 1/2 for the LJ and 5/6 for the Coulomb interactions, as is typically done for the AMBER model~\cite{doi:10.1021/ja00124a002}. Ions placed at a distance of 4 bonds (Hs bonded to the Ns in the FA) feel the whole LJ and Coulomb interactions.

\section{Parameterization of the model}

\subsection{DFT methodology}

\begin{figure}
\includegraphics[width=0.5\textwidth]{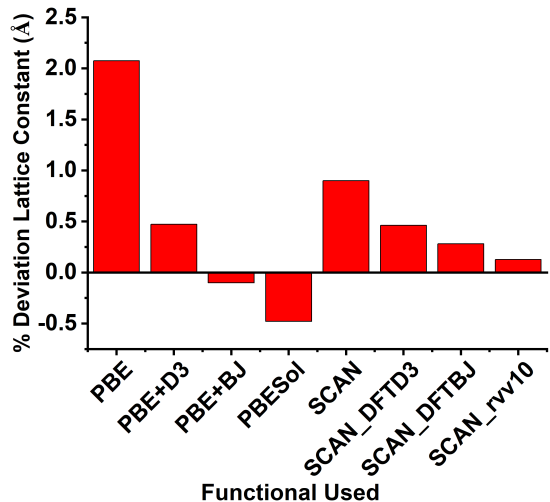}
\caption{Deviations in lattice constant between theory and experiment\cite{weller_cubic_2015} for all functionals tested.}
\label{fig:Bipasa1}
\end{figure}

Our classical FF were trained against a set of reference data calculated with density functional theory (DFT) with the PBE + DFT-D3(BJ) exchange-correlation functional~\cite{PhysRev.136.B864,PhysRev.140.A1133,https://doi.org/10.1002/jcc.21759} in the VASP software package.~\cite{PhysRevB.47.558,KRESSE199615} The choice of DFT-D3(BJ) were made based on extensive comparison of the performance of several functionals, PBE, PBEsol and PBE with vdW corrections, like PBE-D3~\cite{doi:10.1063/1.3382344}, PBE-D3(BJ)~\cite{https://doi.org/10.1002/jcc.21759}, and meta-GGA functional SCAN~\cite{PhysRevLett.115.036402} in combination with several vdW correction schemes, including D3, D3(BJ) and rvv10~\cite{PhysRevX.6.041005}. Our results using pseudo-cubic MAPbI$_3$ or FAPbI$_3$ in Fig.\ref{fig:Bipasa1} indicate PBE-D3(BJ) and SCAN-rvv10 perform the best in predicting the lattice parameters and volume with minimum differences ($<$ 0.1\% and $<$ 0.2\% for lattice parameters and volume) compared to the experiments.\cite{weller_cubic_2015} This highlights the importance of including the vdW corrections when studying hybrid perovskites involving organic cations. Considering SCAN-rvv10 being computational expensive and the need of creating a large training data set, we have chosen PBE-D3(BJ) for the rest of the DFT calculations. The outermost s, p, and d (in the case of Pb) electrons were treated as valence electrons, whose interactions with the remaining ions were modelled by pseudopotentials generated within the projector-augmented wave (PAW).~\cite{PhysRevB.59.1758,PhysRevB.50.17953}.
We have taken the initial bulk structure of cubic FAPbI$_3$ and MAPbI$_3$ from literature~\cite{Tao2017}. For  all  the  calculations  of distorted  structures,  the ions were relaxed while keeping the lattice parameters fixed.  Energy and  force  convergence  threshold  were kept  at  10$^{-4}$ and  -0.05 eV/~\r{A} respectively.  Gaussian smearing with a sigma value of 0.02 is used. An energy cut off of 500 eV and K-points of 6×6×6 are considered for all structures of pseudo-cubic perovskites described below. 

We have modelled electrostatic interactions using atomic point charges. The point charges of all chemical elements in the perovskite materials were calculated using the Density  Derived  Electrostatic  and  Chemical  (DDEC6) method~\cite{doi:10.1021/ct100125x,doi:10.1021/ct3002199}, which was proven to be useful in predicting the electrostatic interactions in halide perovskites.~\cite{li_Tao_fluorides_2019} 

After optimizing the bulk structure, we have generated the distorted structures in various ways to produce a large number of distorted structures to fit the GA (see next section). The following distortions have been used:\cite{D0TA03200J}
\begin{itemize}
\item Contraction and expansion of the length of the unit cell by 0.1~\r{A}. This was applied to all three sides all together and individual sides too while keeping two of the sides intact and varying the other one.
\item Contraction and expansion of the angles of the unit cell by 1 degree. This was applied to all three angles all together and individual angles, keeping two of the angle intact and varying the other one.
\end{itemize}

Since we are dealing with perovskites which have a central organic molecule, we have also oriented them with different angles. We have rotated them along the $a$, $b$ and $c$ directions by 22.5~degree individually till 360~degree to produce a set of distorted structures. 

\begin{figure}
\includegraphics[width=0.48\textwidth]{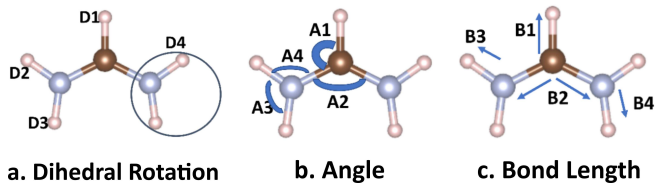}
\caption{Illustration of distortions carried out to calculate the parameters involved in Eq.~\eqref{eq:AMBER}.}
\label{fig:Bipasa2}
\end{figure}

To obtain the parameters of the AMBER model, bonds, angles and dihedrals of the organic molecule were distorted as shown in Fig.~\ref{fig:Bipasa2}. Further details are presented in the next subsection. To obtain the energy change produced by only distorting the organic molecule we have performed single point calculations instead of relaxations. 

\subsection{Fitting bond, bend, and torsion parameters for the organic cations}
In this section we detail how we have proceeded to obtain the force constants, angles and equilibrium distances for the interactions in the molecular mechanics of organic cations. For the bonded terms of MA we took the parameters directly from the GAFF library~\cite{https://doi.org/10.1002/jcc.20035}, because they have proved to reproduce accurately the behavior of the molecule studied here, as have been shown by Mattoni et al~\cite{doi:10.1021/acs.jpcc.6b11298,Mattoni_2016}.

In the case of the FA, as shown in Fig.~\ref{fig:MA_FA_molecules}, there is a double bond, which can lead to confusion when trying to reproduce its effects by a classical FF. One could think in modeling the two Ns in the FA with two different bonds, one with a single bond and the other with a double bond. 
However, it is more realistic to assume that both bonds are equivalent with an intermediate strength between a single and double bond (conjugated bond).
Bearing all this in mind, we include 6 different atoms in the AMBER model as follows: C$_\text{FA}$, C$_\text{MA}$, N$_\text{FA}$, N$_\text{MA}$, H$_\text{N}$, H$_\text{C}$. Added to the Pb, I, Br, we work with 9 different atoms/ions in total.

To obtain values for the FA parameters, we started by taking them from the same library. However, the parameters for the angle interaction between a "n3" nitrogen, a "c2" carbon and a "hx" hydrogen (according to the GAFF notation) are missing. That means that nobody has modeled such interaction before, or if somebody have, it has not been included in the library. In a first try, we used angle parameters of similar elements than the ones included in this case, but the result was that the FAPbI$_3$ compound was unstable when running a MD simulation with our first parameter set. This result was not surprising because, after all, the GAFF parameters have been obtained as averages of AMBER parameters of different molecules with the same kind of elements and bonds, which can provide poor results in a complex $charged$ molecule such as FA. To ensure good reproducibility of the experimental results and the stability of the compounds in the subsequent classical molecular dynamics simulations, we developed a set of new parameters (no GAFF based) for the intramolecular interactions of the FA cation, compatible with the AMBER formulation. We proceeded by fitting the energy of the system to energies obtained by first principles for different distortions of the bonds, angles and dihedrals of the FA molecule in a FAPbI$_3$ unit cell. We assumed that the energy change in the unit cell is entirely due to the internal change produced in the molecule by the distortion, and not to the change in the interaction between the displaced atoms in the molecule with the elements of the crystal lattice because of the small changes in the relative distances between them.

\begin{figure}
\includegraphics[width=0.45\textwidth]{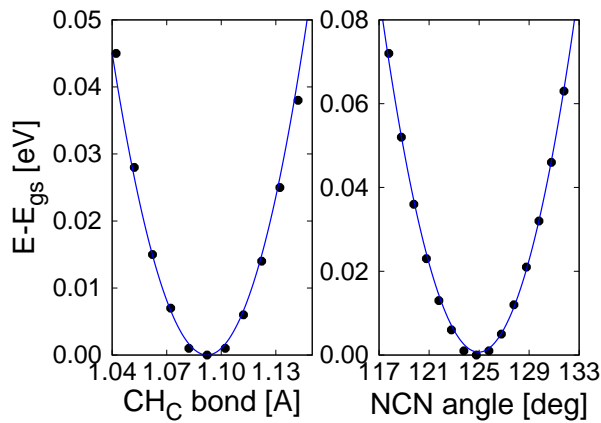}
\caption{Fits to the energy difference between each distortion of the FA and the ground state, calculated by DFT. On the left we show it for the CH$_\text{C}$ bond, and on the right we show it for the NCN angle, as representative cases.}
\label{fig:bond_ang_fit_paper}
\end{figure}

The calculations were performed as follows.
To parameterize the bonds, we manually generated various geometries by stretching and shrinking a chosen bond taking advantage of the symmetry of the molecule. In Fig.~\ref{fig:bond_ang_fit_paper} we present, as an example, the DFT energy difference ($\Delta E=E-E_\text{gs}$) for the C-H$_\text{C}$ bond.
In this case, a simple fit to an harmonic function give us the values of the elastic constant $K$ and the equilibrium distance $r_0$ for this bond.

To parameterize the angles is more tricky, because distorting an angle to check how the energy changes causes indirectly a change in another one. For example, to obtain the parameters for the NCH$_\text{C}$ angles, when increasing the left angle $\theta$ (notation of Fig.~\ref{fig:MA_FA_molecules}), the right angle $\theta '$ decreases. In this specific case, as the angles are equivalent, the elastic constants are the same. Also, the angle differences, as they are squared, will provide the same result. Thus, the energy difference can be obtained as
\begin{align} \label{eq:NCHc_angle}
\Delta E&=K_\text{NCH$_\text{C}$}(\theta -\theta_0)^2 + K_\text{NCH$_\text{C}$}(\theta ' -\theta_0)^2 \nonumber \\ &=2K_\text{NCH$_\text{C}$}(\theta -\theta_0)^2.
\end{align}
This equation can be directly fitted to the DFT calculated reference energies.
To obtain the NCN angle parameters, we distorted the molecule by increasing and decreasing the angle treating the NH$_2$ group as a rigid set. As mentioned before, distorting this angle modifies indirectly the NCH$_\text{C}$ angle, whose parameters have already been calculated by Eq.~\eqref{eq:NCHc_angle} and are necessary to obtain the new ones
\begin{align} \label{eq:NCNangle}
\Delta E=K_\text{NCN}(\theta -\theta_0)^2 +K_\text{NCH$_\text{C}$}(\varphi -\varphi_0)^2 \nonumber \\ \Rightarrow K_\text{NCN}(\theta -\theta_0)^2 = \Delta E -K_\text{NCH$_\text{C}$}(\varphi -\varphi_0)^2. 
\end{align}
By performing another manual fit these parameters can be calculated. This calculation is also presented in Fig.~\ref{fig:bond_ang_fit_paper}, where we can observe that the harmonic function perfectly fits the DFT energy differences. 

To obtain the parameters of the 3 angles that have the N as the central atom, we have performed 3 different sets of calculations: rotating one of the H$_\text{N}$s, the other, and both of them maintaining the angle between them. By doing that we obtain the following set of equations that allow us to determine the rest of the missing parameters
\begin{align}
\Delta E_1 &=K_1(\alpha -\alpha_0)^2 +K_3(\gamma -\gamma_0)^2
\nonumber \\ \Delta E_2 &=K_2(\beta -\beta_0)^2 +K_3(\gamma -\gamma_0)^2
\nonumber \\ \Delta E_3 &=K_1(\alpha -\alpha_0)^2 +K_2(\beta -\beta_0)^2.
\end{align}
where $\alpha$, $\beta$ and $\gamma$ are the bond angles (notation of Fig.\ref{fig:MA_FA_molecules}.)

To obtain the dihedral parameters, because of the complexity of the distortions and the number of possibilities, we just took some examples of distorted structures in which the dihedral contribution to the energy is clearly relevant and performed some fits to obtain a first approximation. The final values were obtained by including the first values as initial guesses in the genetic algorithm. We did that to improve the energy fitting to the DFT energy values and to avoid the complexity of the dihedral calculations.


Final values of the classical FF parameters, including Lennard-Jones, Buckingham and AMBER models for both ion and molecules are presented in the Supporting Information.

\subsection{Electrostatic interactions}
We have put a great effort into developing a classical FF able to efficiently model dynamical properties of the mixed perovskites while ensuring structural stability. Besides, we tried to keep the model as simple as possible to allow for large scale atomistic simulations. For this reason, we assume here some limitations by omitting the possible polarisation of the structural anions I and Br in the precise structural description. Hence, we restricted the model to the derivation of optimized atomic point charges as used in previous classical FFs for perovskite materials.\cite{D0TA03200J,Mattoni_2016,doi:10.1021/acs.jpcc.6b11298} For these point charges, we first developed our model using the values obtained with the DDEC6 method, as described in the DFT methodology section, and we modified them to meet the following requirements:
\begin{itemize}

\item The MA and FA ion molecules, as a whole should have the same charge.
\item The Br and I ions should have the same charge.
\item The system should be electrically neutral.
\item For simplicity, charge values of the H$_\text{C}$s and the H$_\text{N}$s are fixed equal in both molecules.
\end{itemize}
First tries with the GA ({\it vide infra}) using the charges derived in this way did not result in a stable FF for perovskites in the molecular dynamics simulation. Inclusion of an extra parameter in the algorithm, a factor that multiply all the charges at the same time (in order to keep charge neutrality) demonstrated to be really critical, because a small change in it would affect the energy of all the Coulomb interactions of the system, what gives them a predominant role over the rest of the parameters involved in the model. For this reason, and in order to find a set of charges that ensured physical stability of the subsequent molecular dynamics simulation with the classical FF, we modified manually the value of this common factor. 

It is well-known that in hybrid metal halide perovskites~\cite{brivio_relativistic_2014,manser_intriguing_2016}, the frontier orbitals and the electronic band gap are mainly contributed by the Pb and I/Br orbitals rather than by the organic A cations, whose influence is only marginal. It is then reasonable to think that the interaction of the A cations with the PbI$_3$ framework is less covalent and weaker~\cite{gutierrez-sevillano_molecular_2015} and that it has a more ionic character. Within this line of thought, we found that multiplying the charges by a common factor (in particular, 2.5) such that the charges become closer to the nominal ionic charges (for instance 2 por Pb$^{2+}$ and +1 for the whole organic cations) resulted in a stable model in the molecular dynamics simulation. The final values obtained by this ``human learning" procedure are collected in Table~\ref{tab:charges}. It is important to mention that the resulting high charges determine a FF where most of the attractive contribution to the energy is due to the electric charges of ions and counterions, in line with the more ionic character of this interaction. In this respect, we refer the reader to the discussion on the choice of "c" parameters in the Buckingham potential, presented in the Supporting Information.


\begin{table}[h!]
\begin{center}
\begin{tabular}{|c|c|c|c|}
\hline
ion & $q$ ($|e|$) & ion & $q$ ($|e|$) \\ 
\hline 
\hline
C$_\text{FA}$ & 0.9717 & H$_C$ & 0.28 \\ 
\hline 
C$_\text{MA}$ & -0.4422 & H$_N$ & 0.74 \\ 
\hline 
N$_\text{FA}$ & -1.2735 & Pb & 1.9713 \\ 
\hline 
N$_\text{MA}$ & -0.9531 & Br,I & -1.212 \\ 
\hline  
\end{tabular}
\caption{Final values of the charges of the classical force field.}
\label{tab:charges}
\end{center}
\end{table}

\subsection{Buckingham and Lennard-Jones parameters: genetic algorithm}
The Buckingham and Lennard-Jones parameters were obtained after fixing the values of the charges and the parameters of the AMBER model. We have used a new version of the genetic algorithm (GA) previously developed in our group~\cite{D0TA03200J}, adapted to work with LAMMPS instead of GULP to calculate the energies of different structures. LAMMPS has easier compatibility with the AMBER model; moreover it is the code that we will lately use to perform the MD simulations. 

The GA is an artificial intelligence procedure that replicates natural selection. For the ground state of a compound and for its distorted structures, we have obtained the energy of each configuration by DFT calculations. These energy values will be reference data for the GA. The populations in the GA world are different sets of parameters of our classical FF. Each set of parameters will be worse or better "adapted" to reproduce the DFT energies. After each iteration, the best adapted population survives, and new children are created by mixing the genotype of the survivor sets of parameters. Random mutations are also included in the children to increase the variability of the sample, what allows exploring more regions of the configurational space. Once that a given tolerance factor is small enough or a certain number of iterations have been carried out, we obtain the final set of parameters, the one that better reproduces the DFT calculations of all those that have appeared in the GA world. More specific details of the GA can be found in Ref.~\cite{D0TA03200J}. The new version of the code is available online on \url{github.com/salrodgom/flama}.

Instead of running the GA for the 4 pure perovskites we want to describe, we took advantage of the possibility that the ABX$_3$ formula allows to directly mix Br and I. By doing that we are still able to perform calculations with a unit cell of 5 ions (including MA$^{+}$ and FA$^{+}$ ion molecules), and obtain the parameters of the Buckingham potential for the interaction between I and Br atoms directly from the GA. Thus, the two intermediate compounds we have studied are MAPbBr$_2$I and FAPbBr$_2$I. Our reference structure is always the pseudocubic one, because for pure perovskites is the one occurring in solar cells in the operative regime. The exception is the MAPbBr$_2$I, which crystallizes in a tetragonal structure with a very small c/a ratio, so this approximation should be accurate enough. As we have included just one crystal phase of each compound, probably our model potential will not be able to describe phase transitions known to be occurring in these systems. 
However, this is not a fundamental problem since we aim at describing the perovskite material in a solar cell in operation, where phase transitions are neither expected nor desirable. Hence, our classical FF is a ``made-to-purpose'' model and it is not intended to describe the full phase behaviour of the material. 

We have produced and computed around 120 distorted structures for each perovskite. The included distortions can be classified in 3 groups: distortions involving one of the lattice parameters or the three at the same time, distortions involving one of the angles of the unit cell or the three at the same time, and rotations of the organic molecule, avoiding in all cases equivalent situations. This set of distortions gives enough DFT energy values to the GA to be able to fit the model potential with a good level of accuracy. Some of the non-isotropically distorted structures had to be discarded because the energies obtained by DFT calculations were lower than the ground state of the perfect cubic structure. This fact, which is apparently puzzling, is justified by the fact that the cubic structure is not the one with the lowest energy value of the compounds. Some distortions can bring the shape of the cell closer to the phase with the lowest energy value, and provide a lower energy value than the cubic ground state, but a higher energy than the tetragonal phase.

\subsection{Improving transferability}
After running the GA for the pure systems, and in order to have a transferable model between MA and FA, we averaged the common interactions (Pb-Pb, Pb-I,…) obtained in the two different calculations for the two different mixed compounds. We then ran again the GA fixing the common interaction parameters values to let the other parameters readjust to the new averaged values of the common parameters.

To have a complete transferable model, we need to obtain yet the parameters of the interaction between the carbon and the nitrogen atoms of the two different organic molecules. 
In contrast to the case of mixing I and Br, it is not possible to run the GA with a mixture of MA and FA molecules without creating a larger unit cell, because of the stoichiometry of the formula ABX$_3$. Instead, to obtain the missing parameters we averaged the known interactions between C and N atoms that have been obtained in the GA calculations using the following mixing rules
\begin{center}
C$_\text{MA}$ – C$_\text{FA}$ $\rightarrow$ [(C$_\text{MA}$ - C$_\text{MA}$) + (C$_\text{FA}$ - C$_\text{FA}$)]/2 \\
N$_\text{MA}$ – N$_\text{FA}$ $\rightarrow$ [(N$_\text{MA}$ - N$_\text{MA}$) + (N$_\text{FA}$ - N$_\text{FA}$)]/2 \\
C$_\text{MA}$ – N$_\text{FA}$ $\rightarrow$ [(C$_\text{MA}$ - N$_\text{MA}$) + (C$_\text{FA}$ - N$_\text{FA}$)]/2 \\
N$_\text{MA}$ – C$_\text{FA}$ $\rightarrow$ [(N$_\text{MA}$ - C$_\text{MA}$) + (C$_\text{FA}$ - N$_\text{FA}$)]/2
\end{center}

We expect this procedure to work adequately for the following reasons. The interaction between a C$_\text{MA}$ and C$_\text{FA}$ for example is not expected to change significantly with respect to the interaction between two C$_\text{MA}$s or two C$_\text{FA}$s. They are very similar ions, placed at approximately the same distance in both cases, and also surrounded by many H ions, so in the collective interaction of two whole organic molecules these ones will produce just a small amount of the total repulsion felt by the neighbouring atom.

\section{Force field and classical molecular dynamics results}
\begin{figure}
\includegraphics[width=0.45\textwidth]{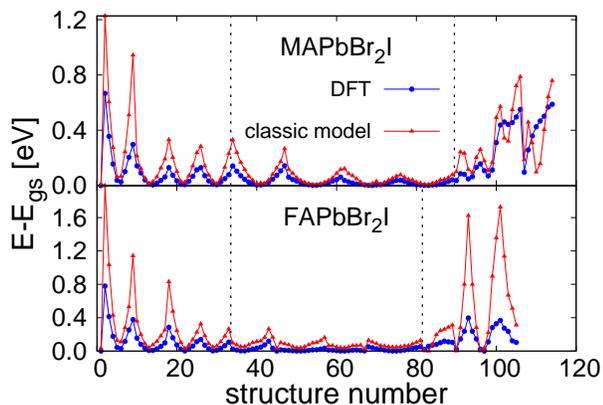}
\caption{Energy difference between each distorted structure and the ground state, for the DFT calculations and for the calculations with the classical FF. We show the results for both of the compounds we have modeled employing the genetic algorithm. The dashed lines distinguish between distortions of the lattice parameter (left), distortions in angles (center) and different orientations of the organic molecule (right).}
\label{fig:DFT_and_model_energies}
\end{figure}

Application of the GA with the procedures described before led to a ``first-guess'' version of the classical FF. In Fig.~\ref{fig:DFT_and_model_energies} we plot the energy of each distorted configuration with respect to the energy of the ground state, from both DFT and from the classical model.
These values are important because they are the ones that appear in the function that the GA uses to establish how well a parameter set of the model reproduces the quantum results (cost function). As can be observed, both lines follow the same trend, except in the distortions corresponding to some orientations of the MA cation, where the energy difference is even higher for some points for the classical FF than for the DFT calculations. We attribute this anomaly to the fact that in the classical calculation there is no contribution of the rearrangement of the electronic structure upon a reorientation of the MA cation. However, in spite of these discrepancies, the behavior of the system is well reproduced, at least qualitatively.

\subsection{Simulation details and stability of the model}
As mentioned before, an important test when developing a FF from scratch is to confirm that the dynamics is stable and that the crystalline structure does not break down. To check this, we ran simulations with a $6\times 6\times 6$ supercell, at 600~K of temperature and 1~atm of pressure for the four pure perovskites. We use a time step of 0.5~fs. Simulations were started from a perfectly cubic unit cell with a lattice parameter of 6.2~\r{A} while forcing the system to stay in a perfect cubic cell (instead of pseudocubic) along the full span of the simulation.  To obtain the volume of the system first we run a 100~ps long simulation in the NPT ensemble and averaged the volume over the last 50~ps. After that, we fixed the volume and thermalized the system in the NVT ensemble for 3~ns. We found necessary to extend the thermalization for such a long time because of the long relaxation times of molecular orientations. A Nosé-Hoover thermostat was used to fix the temperature. Finally, we run a NVE simulation of 20~ns. The energy and the temperature plots versus time show the typical fluctuations in these kind of simulations, which means that the first-guess version of the classical FF is a model with intrinsic stability. Plots of the time evolution of the energy for these exploratory simulations can be found in the Supporting Information (Figure S2).

In spite of the encouraging result regarding stability, we observed that the lattice parameter of the MAPbI$_3$ material as predicted by the simulation (6.0~\r{A}) was too short in comparison with literature values.\cite{doi:10.1021/acs.jpcc.6b11298} Although the rest of the  lattice parameters were in much better agreement~\cite{govinda2018,weller_cubic_2015}, it was found necessary to manually modify the model to correct this anomaly. Hence, we multiplied by 200 the $A$ parameter of the Buckingham potential of the interactions between the C and N of the MA with the I. As this parameter represents the strength of the repulsion between these ions, increasing it leads to a larger volume of the MAPbI$_3$ unit cell, as desired. 

\subsection{Lattice parameters}
\begin{figure}
\includegraphics[width=0.45\textwidth]{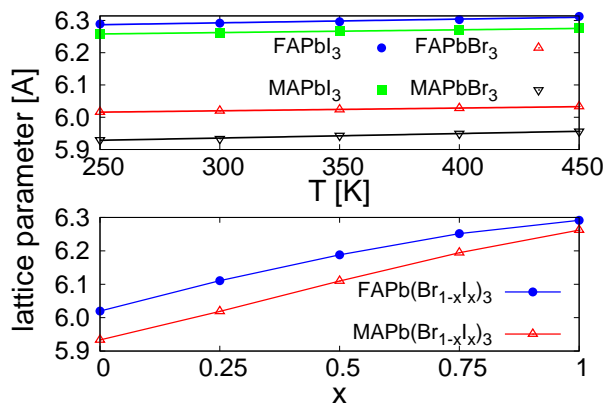}
\caption{Lattice parameters of the four pure compounds we can model, obtained using the final version of the classical FF. We represent them as a function of the temperature (top) and as a function of the I/Br ratio (bottom).}
\label{fig:lattice_parameters}
\end{figure}

The parameters of the final version of the classical FF are presented in the Supporting Information. Lattice parameters, as predicted by the classical FF, are shown in  Fig.~\ref{fig:lattice_parameters} as a function of temperature and composition. 

We observe that the FF replicates the expected behaviour, with the FA perovskites larger than MA ones, and with a systematic expansion of the lattice with respect to temperature and increasing concentration of iodine atoms. In this respect, we calculated the thermal expansion coefficients using the expression
\begin{align} \label{eq:thermal_exp_coef}
\alpha_V=\dfrac{1}{V}\left( \dfrac{\partial V}{\partial T}\right)_P=\dfrac{3}{L}\left( \dfrac{\partial L}{\partial T}\right)_P.
\end{align}
Results can be found in Table~\ref{tab:thermal_exp}.
\begin{table}[h!]
\begin{center}
\begin{tabular}{|c|c|}
\hline
Compound & $\alpha_V$ (K$^{-1}$) \\ 
\hline 
\hline
FAPbI$_3$ & $5.5\times 10^{-5}$ \\ 
\hline 
MAPbI$_3$ & $4.1\times 10^{-5}$ \\
\hline 
FAPbBr$_3$ & $4.2\times 10^{-5}$ \\ 
\hline 
MAPbBr$_3$ & $7.0\times 10^{-5}$ \\
\hline  
\end{tabular}
\caption{Volumetric thermal expansion coefficients of the four pure compounds, calculated by Eq.~\eqref{eq:thermal_exp_coef}.}
\label{tab:thermal_exp}
\end{center}
\end{table}
All of them are of the order of $10^{-5}$~K$^{-1}$ which compare well with an $\alpha_V$ value of $1.57\times 10^{-4}$~K$^{-1}$ for the MAPI~\cite{doi:10.1021/acs.inorgchem.5b01481} and $\alpha_V=2.2\times 10^{-4}$~K$^{-1}$ for the mixture FA$_{0.5}$MA$_{0.5}$PbI$_3$~\cite{C6TA06607K}. However, in the first reference, authors indicate that their value is fairly high, and that other thin-film solar cell materials present an $\alpha_V$ of the order of $10^{-5}$~K$^{-1}$, which would be in better agreement with our results.

\subsection{XRD patterns}
\begin{figure}
\includegraphics[width=0.45\textwidth]{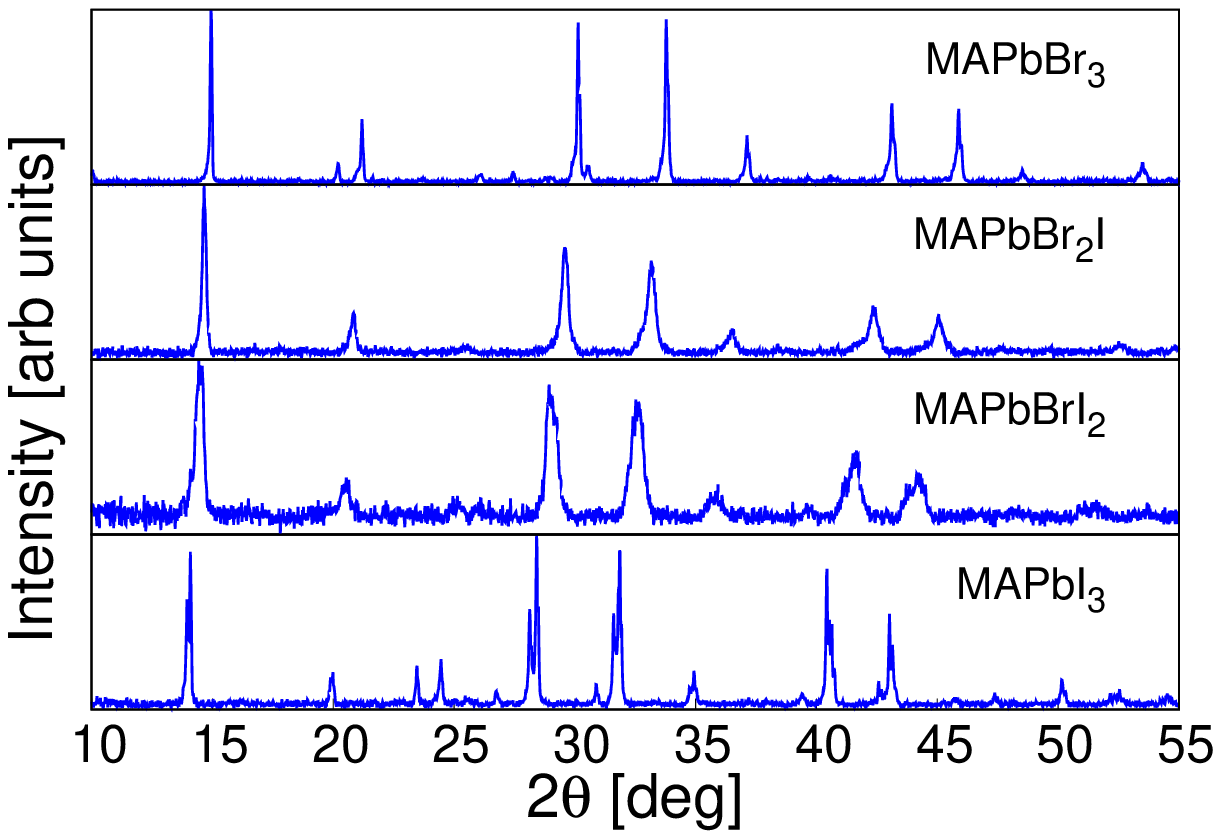}
\includegraphics[width=0.45\textwidth]{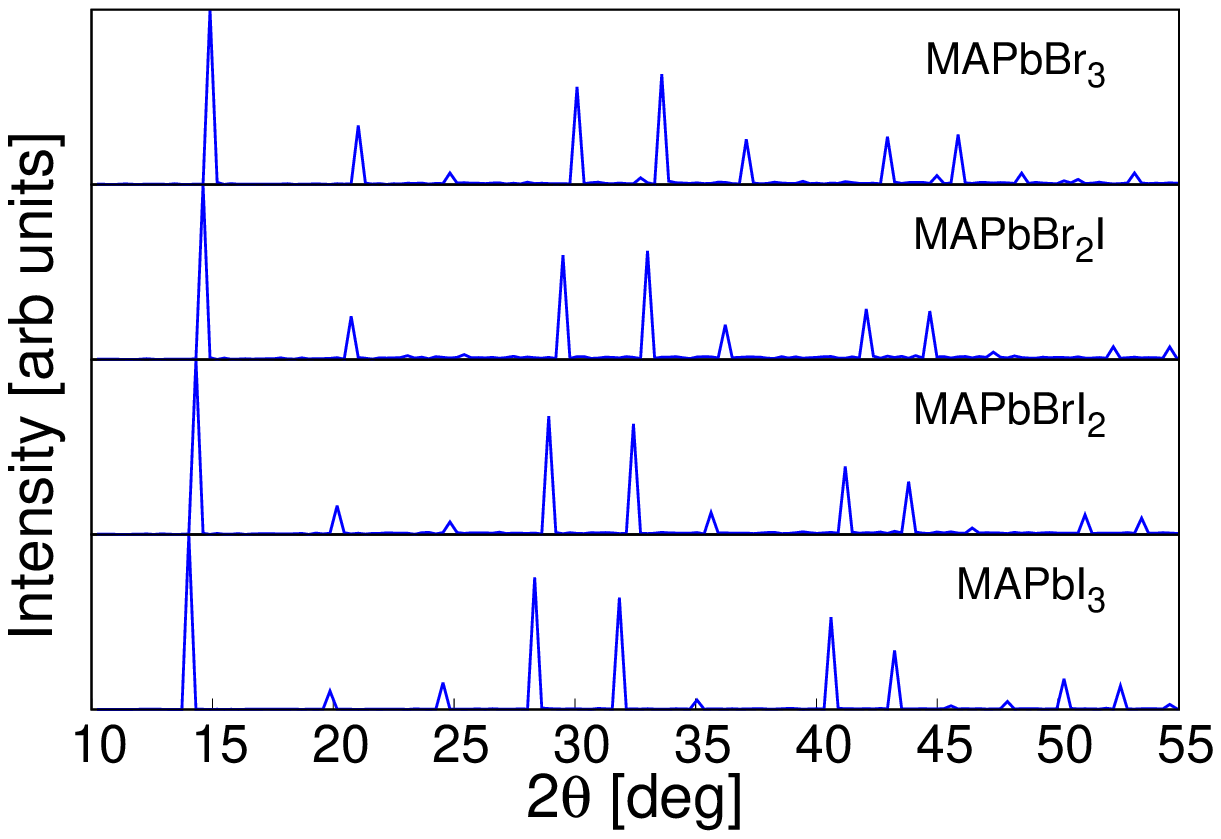}
\caption{Experimental (top) and simulated (bottom) XRD patterns of the MA perovskites.}
\label{fig:XRD_MA}
\end{figure}

\begin{figure}
\includegraphics[width=0.48\textwidth]{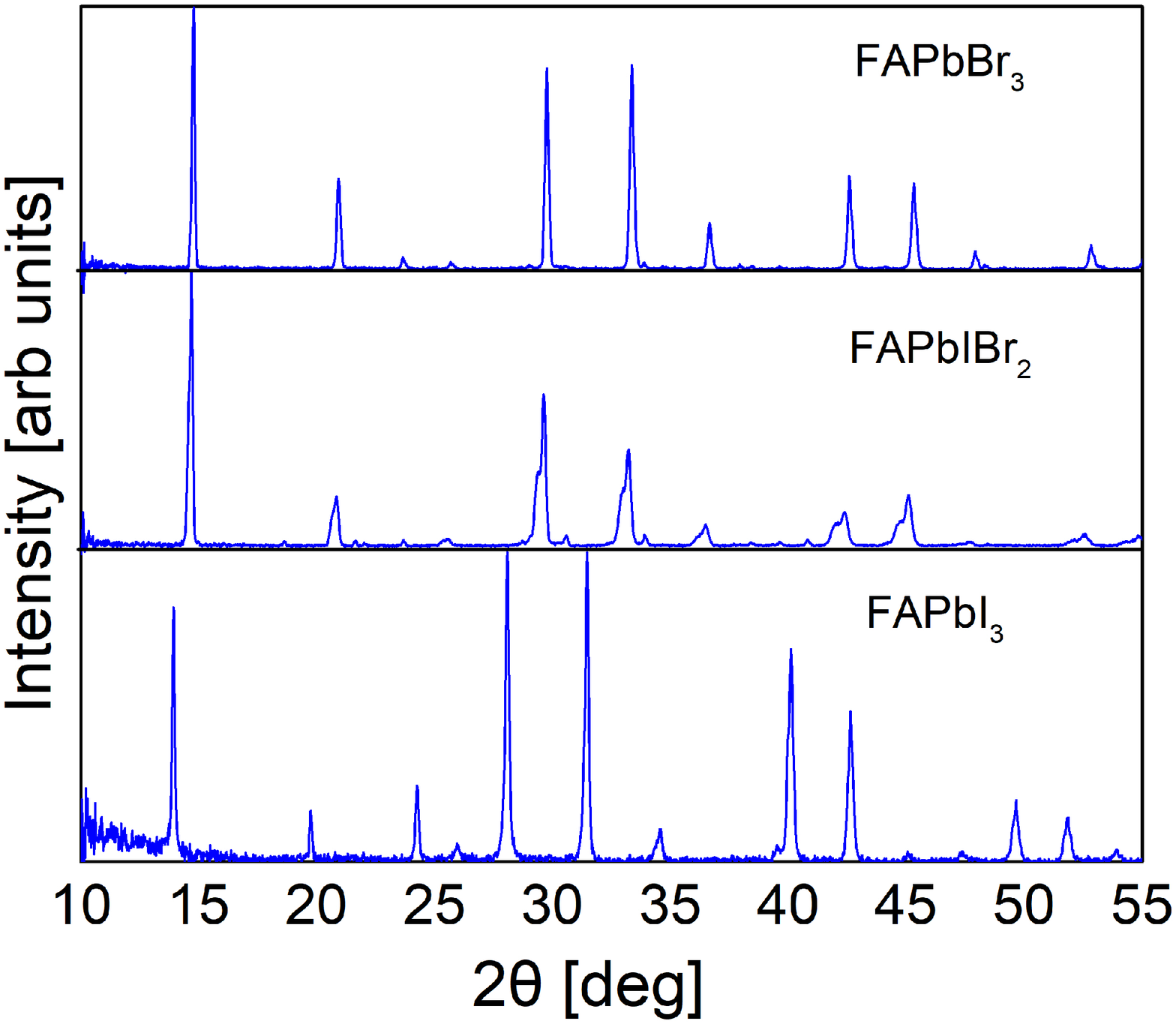}
\includegraphics[width=0.39\textwidth]{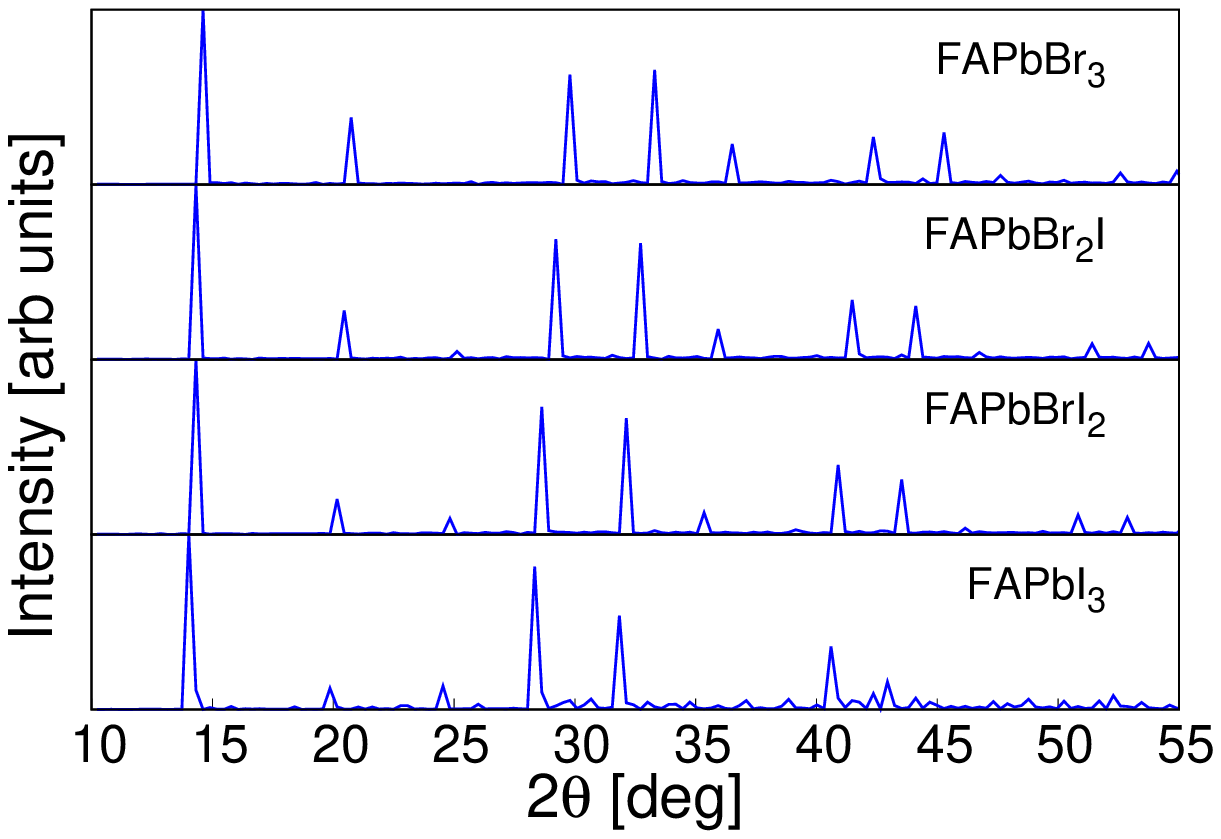}
\caption{Experimental (top) and simulated (bottom) XRD patterns of the FA perovskites.}
\label{fig:XRD_FA}
\end{figure}

We have checked how well the classical FF reproduces the experimental XRD patterns of the studied perovskites. Simulated and experimental results are compared in Fig.~\ref{fig:XRD_MA} for MA perovskites and in Fig.~\ref{fig:XRD_FA} for FA perovskites. Experimental details about the preparation and XRD characterization of perovskite films and powders are collected in the Supporting Information.

The substitution of the iodide ion for the smaller bromide ion in both FA and MA perovskite systems is followed by a shift toward higher angles in the positions of the XRD pattern peaks because of the corresponding reduction in interplanar distances. This shift is accurately reproduced by CMD simulations with the GA-fitted classical FF. In general, XRD patterns are well replicated by the model with a cubic phase, in spite of the fact that in the experiment the perovskite is in a tetragonal phase with small $c/a$ ratio. However, the similar shift produced by FA/MA substitution in the $\alpha$-black phase of FA perovskites is not completely caught by the model although this deviation is actually very small (see Supporting Information,  Figure S3).

\section{Dynamical properties}

One of the main objectives of our made-to-purpose model is to reproduce correctly the dynamical properties and the ion migration features of the mixed perovskites at typical working conditions of the solar cell. Specifically, we have run simulations of 10~ns in the same conditions we used to check the stability of the model to obtain the diffusion coefficients of the different ions. To achieve that, we have calculated the mean squared displacement (MSD) of each one of them. MSD is defined as a statistical average of the change in the ion positions in the simulation
\begin{align}
\text{MSD}=\left< (\vec{r}(t)-\vec{r}(0))^2\right> = \dfrac{1}{N}\sum_{i=1}^N \parallel \vec{r}(t)-\vec{r}(0) \parallel^2
\end{align}
In this equation, the index $i$ runs over all the $N$ migrating ions. If the ions follow a diffusive regime, the MSD would increase linearly with time, and will be related with the diffusion coefficient as
\begin{align}
D=\dfrac{1}{6t}\text{MSD}.
\end{align}

Results for typical MSDs can be found in the Supporting Information in Figure S4. Within this formalism, we have not been able to appreciate diffusion of neither Pb ions nor the organic molecules, even using a large number of vacancies and a temperature of 600~K. This is sort of expected knowing the activation energies of the Pb and the MA (2.31 and 0.84~eV respectively) and also the reported diffusion coefficient of the MA ($10^{-16}$~cm$^2$~s$^{-1}$)~\cite{Eames2015}. However, we succeeded in obtaining the diffusion coefficients of the halide ions. For each pure compound, we ran simulations for five temperature values and for two alternative number of vacancies. Moreover, for each combination of these two variables, we run three simulations with different initial conditions, locating the vacancies in different positions, to improve the statistics and reduce the numerical error.

\begin{figure}
\includegraphics[width=0.45\textwidth]{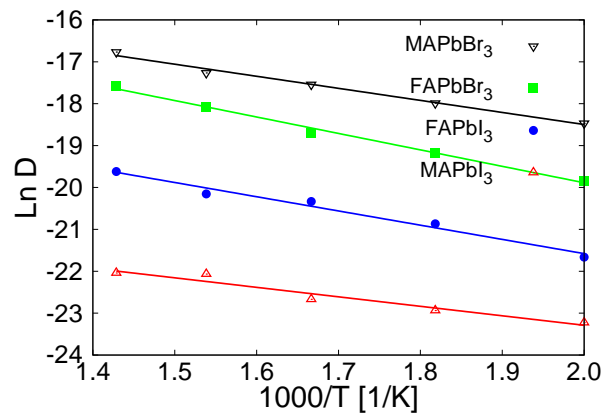}
\caption{Fit to the Arrhenius equation (Eq.~\eqref{eq:Arrhenius}) for the four pure perovskites in simulations with 24 vacancies. Each point in this plot is the average of the results obtained in 3 simulations with different initial conditions.}
\label{fig:LnD_vs_1000T}
\end{figure}

It is important to mention that at 300~K diffusion is so slow that the required simulation time exceeds common computational resources. So we decided to calculate the diffusion coefficient at higher temperatures and extrapolate it to room temperature using Arrhenius equation
\begin{align} \label{eq:Arrhenius}
D=D_0\exp \left(-\frac{E_a}{RT}\right),
\end{align}
where $D_0$ is a temperature-independent preexponential factor, $E_a$ is the activation energy for diffusion, and $R$ is the gas constant.
In Fig.~\ref{fig:LnD_vs_1000T} we plot the fit to Eq.~\eqref{eq:Arrhenius} for the four pure perovskites, from simulations run with 24~vacancies (approx. 5 $10^{20}$ cm$^{-3}$). Each point is an average over the results of three different initial conditions.
It can be seen that Arrhenius equation fits accurately the simulated data. From the fitting formula we can extrapolate the diffusion coefficients at 300~K and calculate the activation energy of each species. The results are presented in Table~\ref{tab:D_Ea}.

We find values of the order of $10^{-10}-10^{-11} $cm$^2$/s for the diffusion coefficients and activation energies between 0.2 and 0.33 eV. We also observe that values from the simulations with 24 vacancies are more or less twice the ones of the simulations with 12 vacancies, as predicted by the theory ({\it vide infra}). Moreover, the activation energies of the studied systems do not depend on the number of vacancies we have included in the simulation, and their values are in good agreement with the energies reported in the literature.\cite{Eames2015,Meggiolaro2018} For MAPbI$_3$ diffusion is so slow that no diffusion coefficient could be derived from a simulation with 12 vacancies of just 10~ns.
\begin{table}[h!]
\begin{center}
\begin{tabular}{|c|c|c|c|}
    \hline
    Compound & N vac & $D_\text{300 K}$~[cm$^2$ s$^{-1}$] & $E_a$ [eV/ion] \\ 
    \hline
    \hline
    \multirow{2}{*}{MAPbBr$_3$}&12&1.4 $\times$ 10$^{-10}$&0.27\\
    &24&5.2 $\times$ 10$^{-10}$&0.25\\
    \hline
    \multirow{2}{*}{FAPbBr$_3$}&12&1.6 $\times$ 10$^{-11}$&0.32\\
    &24&4.6 $\times$ 10$^{-11}$&0.34\\
    \hline
    \multirow{2}{*}{MAPbI$_3$}&12&-&-\\
    &24&8.0 $\times$ 10$^{-12}$&0.20\\
    \hline
    \multirow{2}{*}{FAPbI$_3$}&12&5.0 $\times$ 10$^{-12}$&0.29\\
    &24&1.4 $\times$ 10$^{-11}$&0.29\\
    \hline
\end{tabular}
 \caption{Diffusion coefficient at 300~K and activation energy for each pure perovskite with a two concentrations of vacancies in the supercell.}
\label{tab:D_Ea}
\end{center}
\end{table}

From the results obtained with two concentrations of vacancies, we can infer the jumping rate, which is independent of the vacancy concentration, and use it to extrapolate the diffusion coefficient to a realistic density of vacancies at room temperature. To do so, we have applied the following procedure, described in references~\cite{https://doi.org/10.1002/adfm.201500827,C8EE01697F}. The diffusion coefficient $D_i$ of the specie $i$ (which for us will always be I or Br) is related to the jumping rate of the same species $\Gamma_i$ by
\begin{align} \label{eq:D_i_1}
D_i=\dfrac{1}{6}Zd^2\Gamma_i ,
\end{align}
where $Z$ is the number of neighbours (8 in our case) and $d$ is the jumping distance. In our case, as we are considering cubic cells and jumps from one site of the octahedra to another, we have
\begin{align}
d^2=\left( \dfrac{a}{2} \right)^2+\left( \dfrac{a}{2} \right)^2=\dfrac{1}{2}a^2,
\end{align}
where $a$ is the lattice parameter. From Eq.~\eqref{eq:D_i_1} we calculate the jumping rate, that is also related with the jumping rate of the vacancies $\Gamma_v$ of that species by
\begin{align}
\Gamma_i=\dfrac{N_v}{N_i}\Gamma_v,
\end{align}
where the subindex $v$ is used for vacancies, and $N_v$ and $N_i$ are the vacancies and ions concentrations, respectively. In fact, as concentrations always appear related in a fraction, we can directly use the total number of vacancies and ions, what is easier because we fix these values when creating the supercell of the simulation. Once that we know $\Gamma_v$, we can calculate the diffusion coefficient at 300~K with the number of vacancies expected at room temperature via
\begin{align} \label{eq:D_final}
D_i=\dfrac{1}{6}Zd^2\dfrac{N_v}{N_i}\Gamma_v. 
\end{align}

\begin{figure*}[!t]
\includegraphics[width=\textwidth]{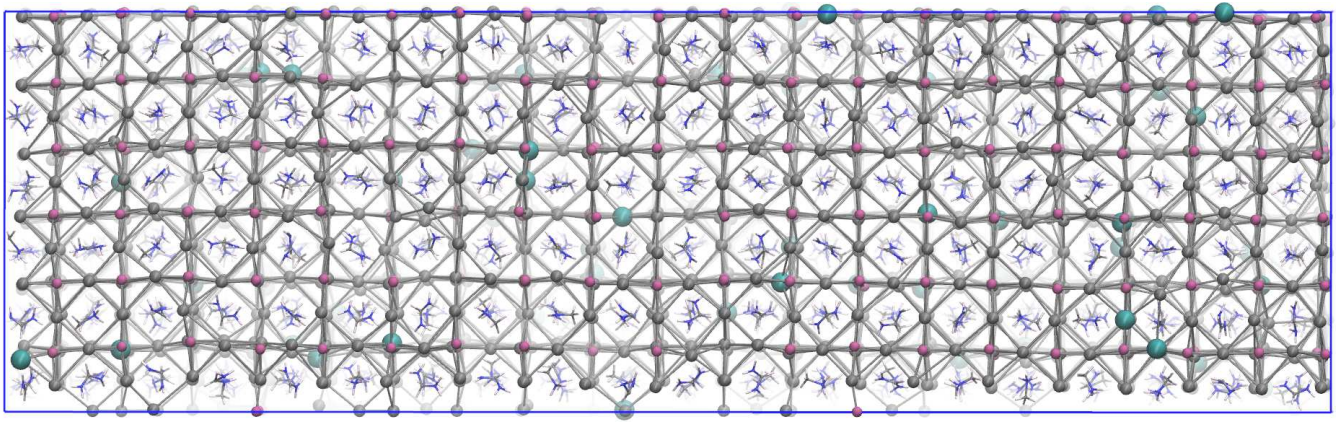}
\caption{\label{fig:supercell}Snapshot of the $6\times 6\times 20$ structure. Code colour of Pb, I, Br, C, N, H atoms: mauve, silver, seawater (larger spheres), grey, blue, and white, respectively.}
\end{figure*}

We have calculated the expected density of vacancies taking into account the reaction enthalpies reported in Ref.~\cite{https://doi.org/10.1002/anie.201409740}. In that work, authors present three different reactions in which one or more I vacancies in the MAPbI$_3$ perovskite are involved, which is the defect that enables diffusion. We expect these reactions to provide a more realistic vacancy concentration than other calculations that uses DFT to obtain the energy formation of a defect in isolation~\cite{doi:10.1021/acsenergylett.9b00247}. Taking into account the three reactions studied in Ref.~\cite{https://doi.org/10.1002/anie.201409740}, the 4.25\% of the I sites would be empty, which corresponds to a vacancy concentration of the order of $10^{20}$~cm$^{-3}$ for MAPbI$_3$. In any case, our procedure makes it possible to estimate the actual diffusion coefficient for any temperature and vacancy concentration, should this be required for different experimental conditions or alternative arguments by which a different vacancy concentration can be argued.~\cite{bertoluzzi2019,weber2018} 

As the volume of the unit cell of all the studied perovskites is different, we find more sensible to extrapolate the result of 4.25\% empty halide locations rather than the vacancy concentration to estimate this parameter in all perovskites. Hence. using Eq.~\eqref{eq:D_final} we are in the position to estimate the diffusion coefficient of the halides in the four pure perovskites at a temperature of 300~K for a realistic vacancy concentration. The results are presented in Table~\ref{tab:final_pure_results}. 
\begin{table}[h!]
\begin{center}
\begin{tabular}{|c|c|}
\hline
Perovskite & $D_\text{300 K}$~[cm$^2$ s$^{-1}$] \\ 
\hline 
\hline
MAPbBr$_3$ & 4.6 $\times$ 10$^{-10}$ \\ 
\hline 
FAPbBr$_3$ & 4.5 $\times$ 10$^{-11}$ \\ 
\hline 
MAPbI$_3$ & 9.2 $\times$ 10$^{-12}$ \\ 
\hline 
FAPbI$_3$ & 1.4 $\times$ 10$^{-11}$ \\ 
\hline  
\end{tabular}
\caption{Predicted values of the diffusion coefficient of  halides in the four pure compounds with pseudocubic structure at 300~K and a vacancy concentration corresponding to 4.25\% of empty halide sites (approx. $10^{20}$~cm$^{-3}$).}
\label{tab:final_pure_results}
\end{center}
\end{table}

As inferred from the results, we predict a faster diffusion for the Br ions than for the I. This is an important result because it would be a factor for intrinsic instability and phase segregation of perovskites in operation.\cite{li_phase_2017,brennan_photoinduced_2020} 

It is important to mention that our results are one order of magnitude higher than some results reported in the literature~\cite{D0NR03058A,Eames2015}. For instance, in order to reproduce the hysteresis of the IV curve for typical scan rates of 100 mV/s and a (thermally activated) frequency peak of 0.1 -1 Hz in the impedance spectra, a diffusion coefficient of the order of 10$^{-12}$ cm$^2$/s is required for MAPbI$_3$ cells~\cite{Eames2015}. However, we must bear in mind that the values obtained with the procedure we devised here correspond actually to a perfectly cubic crystalline solid, with halide vacancies but without other defects such as grain boundaries. This other type of defects, commonly present in solution processed perovskites, would significantly delay the ion dynamics in solar cells\cite{Phung2020}. The values here reported would then correspond to the theoretical upper limit for ion diffusion in perovskite crystals, as predicted using exclusively first-principles arguments. 

We have also run simulations for the mixtures presented in Table~\ref{tab:mixtures}. In this case, the linearity predicted by Eq.~\eqref{eq:D_final} with respect to the concentration of vacancies is also approximately maintained, and the activation energies are of the same order as for the pure compounds. For perovskites that include I and Br ions, we have computed a combined diffusion coefficient for both of them. We observe that the diffusion increases with increasing values of the Br/I ratio. For the mixed MA-FA case with, we predict an easier mobility for the Br ions when they are in a MA environment with respect to a FA one. We would expect the same behavior when working with I instead of Br. 

\begin{table}[h!]
\begin{center}
\begin{tabular}{|c|c|c|c|}
    \hline
    Compound & N vac & $D_\text{300 K}$~[cm$^2$ s$^{-1}$] & $E_a$ [eV/ion] \\ 
    \hline
    \hline
    \multirow{2}{*}{FAPbBr$_1$I$_2$}&12&1.0 $\times$ 10$^{-11}$&0.27\\
    &24&1.4 $\times$ 10$^{-11}$&0.31\\
    \hline
    \multirow{2}{*}{FAPbBr$_2$I$_1$}&12&1.1 $\times$ 10$^{-11}$&0.32\\
    &24&2.1 $\times$ 10$^{-11}$&0.33\\
    \hline
    \hline
    \multirow{2}{*}{MA$_{0.5}$FA$_{0.5}$PbBr$_3$}&12&3.6 $\times$ 10$^{-11}$&0.29\\
    &24&6.2 $\times$ 10$^{-11}$&0.32\\
    \hline
\end{tabular}
\caption{Predicted diffusion coefficients for various combinations of the FA compounds and the MA-FA mixture.}
\label{tab:mixtures}
\end{center}
\end{table}

In order to rule out system size effects, and to get closer to the macroscospic limit, we have extended the simulations to a $6\times 6\times 20$ supercell ($\sim$8600 atoms) and a total simulation time of 40 ns. These longer simulations were run on a composition of MA$_{0.29}$FA$_{0.71}$PbBr$_{0.1}$I$_{2.9}$~\cite{CASTROCHONG2020836} and with 4.25\% empty halide sites (see Fig.~\ref{fig:supercell}). In these simulations an external electric field up to 0.01~V~\r{A}$^{-1}$ along the $z$~direction have been applied. Ionic segregation was not observed to occur in the system, probably because it is a process that takes place in longer time scales~\cite{D0NR03058A}. However, we have confirmed that larger samples produced statistically equal values of the ion diffusion coefficients than in the simulations with smaller samples, shorter simulations and without electric field, showing that the electric field does not affect the ion migration properties at least in the short time scale. 

\section{Conclusions}

We have derived a classical force field with transferable properties suitable to describe mixed perovskites (MA$_x$FA$_{1-x}$Pb(Br$_y$I$_{1-y}$)$_3$ $\forall x,y \in [0,1]$) by molecular dynamics with up to 8600 atoms and simulation times of up to 40 ns. The force field parameters were determined by means of a genetic algorithm that seeks the best set that fits DFT energies for a combination of crystalline structures with several degrees of distortion with respect to the equilibrium configuration, and then refined to ensure stability of the molecular dynamics simulations. The classical model reproduces correctly the crystalline structure (as inferred from the comparison of the simulated XRD patterns with experimental ones), the expansion of the structure upon Br/I substitution and the thermal expansion coefficients. It also predicts activation energies for halide diffusion in agreement with literature values. Using theoretical arguments based on jumping rates and an extrapolation scheme based on Arrhenius equation, we have been able to predict the ion diffusion coefficients of mixed perovskites at room temperature and for realistic concentrations of ion vacancies. Being based on long simulations and big samples for a DFT-fitted classical forced field, these values can be considered as the best values obtainable from first principles for perfectly crystalline solids with no defects other than halide vacancies. As real perovskites sinthesized for solar cell operation do always have a variable amount of defects and grain boundaries, the value predicted in this work can be considered as a theoretical upper limit for ion diffusion in real perovskites for photovoltaic applications.


%
%
%

\section{Acknowledgements}
This work was funded by the Ministerio de Ciencia e Innovación of Spain, Agencia Estatal
de Investigación (AEI) and EU (FEDER) under grants PID2019-110430GB-C22 and
PCI2019-111839-2 (SCALEUP) and Junta de Andalucia under grant SOLARFORCE (UPO-1259175). SRGB was supported by grant FJC2018-035697-I funded by MCIN/AEI/10.13039/501100011033.
S. T. and J.M.V.L. acknowledge funding support from NWO START-UP from the Netherlands. We also thank C3UPO for the HPC support.

\bibliography{biblio}
\bibliographystyle{hunsrtnat}
\end{document}